\documentclass[sigconf]{acmart}

\AtBeginDocument{%
  \providecommand\BibTeX{{%
    \normalfont B\kern-0.5em{\scshape i\kern-0.25em b}\kern-0.8em\TeX}}}

\usepackage{epsfig}
\usepackage{bm}
\usepackage{url}

\usepackage[T1]{fontenc}
\usepackage{aecompl}

\setcopyright{none}

\makeatletter
\g@addto@macro\normalsize{%
\abovedisplayskip 3pt plus 1pt minus 1pt%
\belowdisplayskip \abovedisplayskip
\abovedisplayshortskip 3pt plus 1pt minus 1pt%
\belowdisplayshortskip 3pt plus 1pt minus 1pt%
}
\makeatother

\copyrightyear{2020}
\acmYear{2020}
\setcopyright{acmcopyright}\acmConference[CIKM '20]{Proceedings of the 29th ACM International Conference on Information and Knowledge Management}{October 19--23, 2020}{Virtual Event, Ireland}
\acmBooktitle{Proceedings of the 29th ACM International Conference on Information and Knowledge Management (CIKM '20), October 19--23, 2020, Virtual Event, Ireland}
\acmPrice{15.00}
\acmDOI{10.1145/3340531.3412700}
\acmISBN{978-1-4503-6859-9/20/10}
\begin{document}

\fancyhead{}
\title{EdgeRec: Recommender System on Edge in Mobile Taobao}

\author{Yu Gong$^{1\ast}$, Ziwen Jiang$^{1\ast}$, Yufei Feng$^{1}$, Binbin Hu$^{2}$, Kaiqi Zhao$^{3}$, Qingwen Liu$^{1}$, Wenwu Ou$^{1}$}
\thanks{$^{\ast}$ Yu Gong and Ziwen Jiang contribute equally.}
\affiliation{
 \institution{$^{1}$Alibaba Group, $^{2}$Ant Financial Services Group, $^{3}$University of Auckland}
}
\email{{gongyu.gy, ziwen.jzw, xiangsheng.lqw, santong.oww}@alibaba-inc.com,
fyf649435349@gmail.com, bin.hbb@antfin.com}

\renewcommand{\authors}{Yu Gong, Ziwen Jiang, Yufei Feng, Binbin Hu, Kaiqi Zhao, Qingwen Liu, Wenwu Ou}


\begin{abstract}
Recommender system (RS) has become a crucial module in most web-scale applications. 
Recently, most RSs are in the waterfall form based on the cloud-to-edge framework, where recommended results are transmitted to edge (e.g., user mobile) by computing in advance in the cloud server. Despite effectiveness, network bandwidth and latency between cloud server and edge may cause the delay for system feedback and user perception. Hence, real-time computing on edge could help capture user preferences more preciously and thus make more satisfactory recommendations.
Our work, to our best knowledge, is the first attempt to design and implement the novel Recommender System on Edge (EdgeRec), which achieves \emph{Real-time User Perception} and \emph{Real-time System Feedback}.
Moreover, we propose \emph{Heterogeneous User Behavior Sequence Modeling} and \emph{Context-aware Reranking with Behavior Attention Networks} to capture user's diverse interests and adjust recommendation results accordingly.
Experimental results on both the offline evaluation and online performance in Taobao home-page feeds demonstrate the effectiveness of EdgeRec.
\end{abstract}


\maketitle

\section{Introduction}
\label{sec:intro}
\begin{figure}[th]
	\centering
	\includegraphics[angle=0, width=1.0\columnwidth]{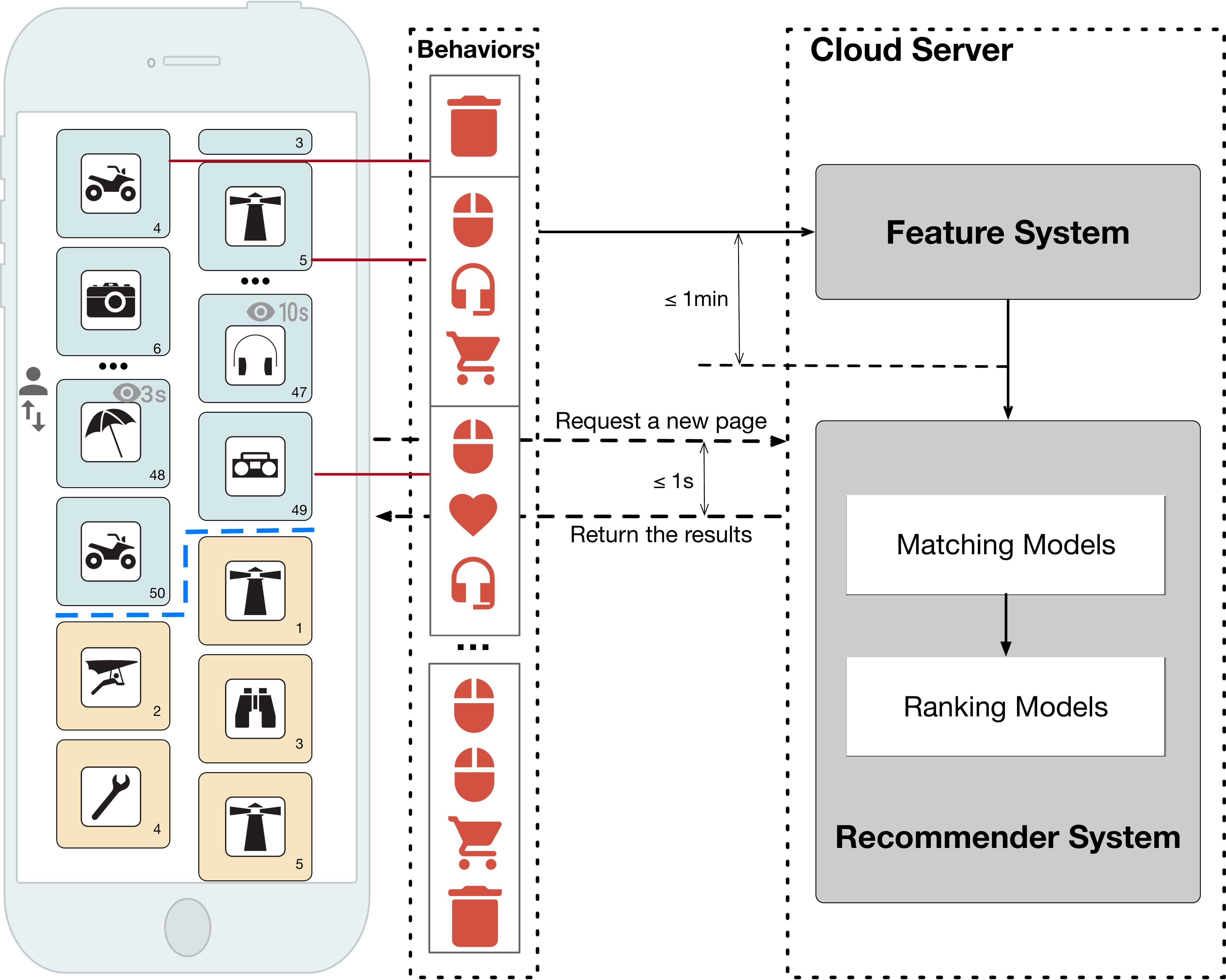}
	\caption{Illustration of the popular cloud-to-edge based waterfall recommender system.}
	\label{fig:motivation}
\end{figure}

The explosive growth and variety of information (e.g., movies, commodities, news, etc.) available on web frequently overwhelm users.
Recommender system (RS) serves as a valuable means to cope with the information overload problem, which selects a list of items from the overwhelmed candidates to meet user's diverse demands.
In most scenarios of commercial RS, especially on mobiles, 
recommended items are shown in a waterfall form. 

As shown in Fig. \ref{fig:motivation}, most waterfall RS are deployed based on the cloud-to-edge framework.
Mobile client first initiates a paging request to the cloud server when a user scrolls on a waterfall RS scenario.
Then matching and ranking models serving on cloud server respond to the paging request and generate a list of ranked items which will be displayed to the user. Under such circumstance, current cloud-to-edge based waterfall RS suffers from the following limitations:
\begin{itemize}
\item \textbf{Delay for System Feedback}: Due to the paging mechanism in the cloud-to-edge framework, RS on cloud is unable to adjust the recommended results in time between the two adjacent paging requests, which further fails to satisfy user's changing demands. Take an example in Fig. \ref{fig:motivation}, the user clicks on a dress in the 5-th position of current page, which reflects his/her sudden preference to the dress category. Nevertheless, RS on cloud cannot respond to it unless the user scrolls into the next page, which consequently fails to satisfy his/her demand in time and decreases the user experience.
\item \textbf{Delay for User Perception}: For RS models serving on cloud, 
up to 1-minute delay is existed
for capturing user behaviors due to network latency, so they fail to model user's real-time preferences when responding to edge.
For example, in Fig. \ref{fig:motivation}, a user's behaviors on the item in the 49-th position reveal his/her preference to radios at present but RS on cloud cannot recommend similar radios in the next page because of not receiving those behaviors in time. 
Moreover, network bandwidth further limits current RS to capture diverse and detailed user behaviors on edge.
\end{itemize}
To summarize, the limitation of RS on cloud is that the delayed adjustment of recommended results cannot match the real-time changes of user's preferences on edge, so as to seriously harm the user experience of commercial RS.

Edge computing \citeN{shi2016edge, vallati2016mobile, taleb2017mobile, CSun} is ideal for applications that require high real-time performance, and has the potential to address the above concerns of current RS based on the cloud-to-edge framework.
Our work takes the first place to design and implement the novel recommender system on edge \footnote{Here the user's mobile device is the edge.} (EdgeRec), which achieves \textbf{Real-time User Perception} and \textbf{Real-time System Feedback} without additional requests to cloud servers.
Our main contributions can be summarized as follows:
\begin{itemize}
\item \textbf{System Architecture} - We design the architecture of EdgeRec to conduct reranking on the mobile device, cooperating with RS on cloud that provides candidate items (see Sec. \ref{sec:system_overview}).
\item \textbf{System Implementation} - EdgeRec supports large-scale neural network models, considering efficient computation and storage on mobile device by distributing the model across edge and cloud (see Sec. \ref{sec:system_implementation}).
\item \textbf{User Behavior Modeling} - We propose \emph{Heterogeneous User Behavior Sequence Modeling} to capture user's changing behaviors and actions. 
We first design the novel feature system (see Sec. \ref{sec:feature}) then simultaneously model positive and negative feedback between users and items considering both interacted items and their corresponding actions (see Sec. \ref{sec:user_state}).
Based on EdgeRec, diverse and detailed user behaviors in our feature system are collected, stored and consumed on edge, which can be fed into our model in real time.
\item \textbf{Context-aware Reranking} - We propose \emph{Context-aware Reranking with Behavior Attention Network} for reranking on edge. Specifically, we model interactions between candidate items and real-time user behavior context by our proposed \emph{Behavior Attention} mechanism (see Sec. \ref{sec:rerank}).
Depending on the ability to rerank items on edge based on EdgeRec, we achieve the real-time response to meet user's demands.
\end{itemize}

We conduct extensive offline and online evaluations on the real traffic of Taobao home-page feeds.
Both quantitative and qualitative analyses demonstrate the rationality and effectiveness of our proposed EdgeRec system.
Furthermore, EdgeRec contributes up to \textbf{1.57\%} PV, \textbf{7.18\%} CTR, \textbf{8.87\%} CLICK and \textbf{10.92\%} GMV promotions in online A/B testing \footnote{PV and CLICK are defined as the total number of items viewed and clicked by the users.
CTR is the click-through rate and is calculated by $\text{CLICK}/\text{PV}$.
GMV is the total amount of money (revenue) user spent on the recommended items.},
which brings significant improvements to current Taobao RS.
Now EdgeRec has already been deployed online and serves the
main traffic.
\section{System}
\label{sec:system}
In this section, we present the EdgeRec system, which aims to capture timely rich user behaviors (a.k.a. Real-time Perception) and respond to users' demands in time (a.k.a. Real-time Feedback) without any additional request to the cloud servers. We start with an overview of the EdgeRec system, followed by the elaboration on implementation for each well-designed module.

\begin{figure}[th]
	\centering
	\includegraphics[angle=0, width=1.0\columnwidth]{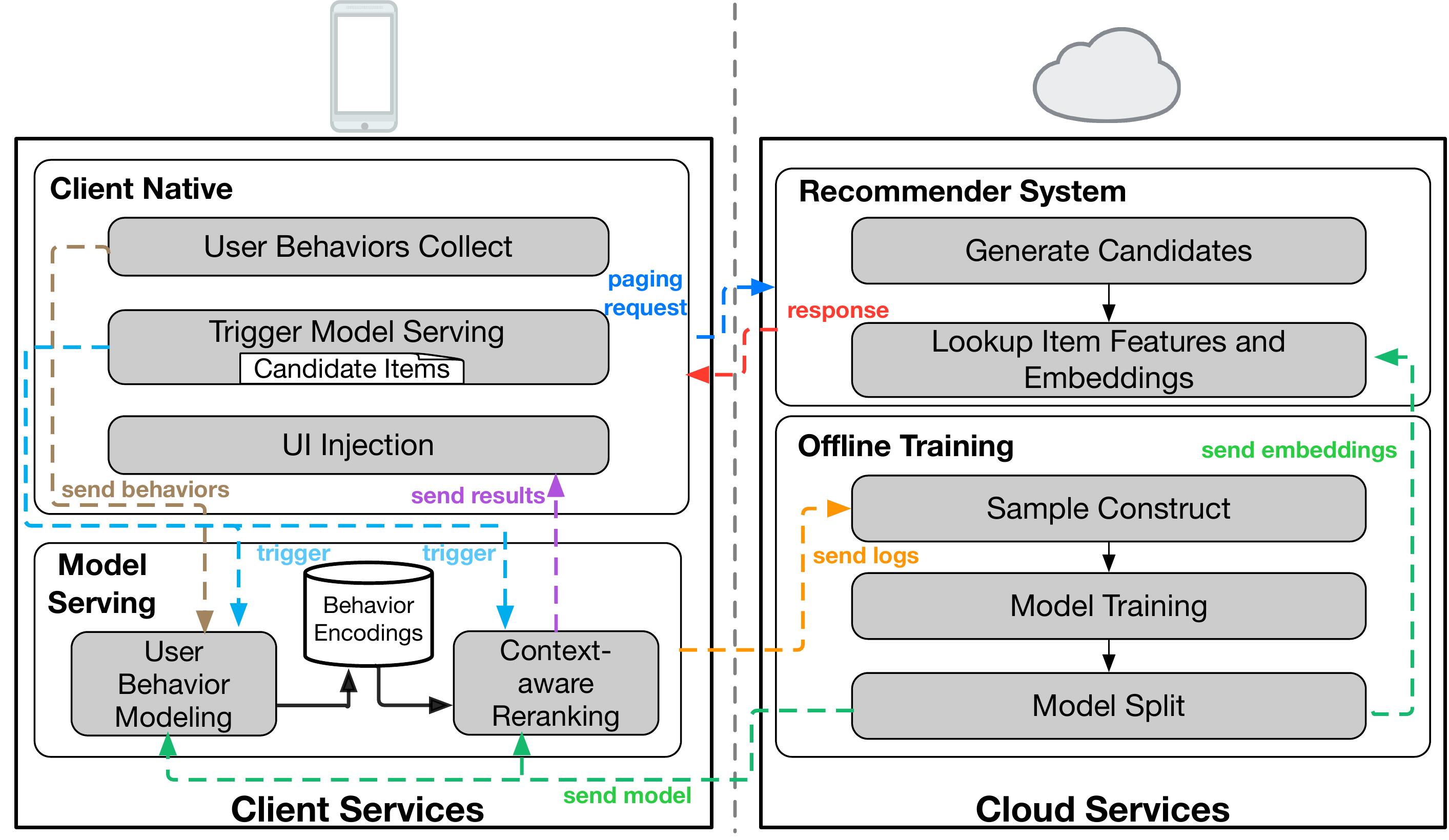}
	\caption{EdgeRec system overview.
		The left part of modules are deployed in mobile Taobao client and the right part of modules are serving on cloud.
	}
	\label{fig:system}
\end{figure}

\subsection{System Overview}
\label{sec:system_overview}
In Fig. \ref{fig:system}, we show the overview of EdgeRec system.
Note that EdgeRec aims to cooperate with RS on cloud instead of replacing it.
The main modules and workflows are illustrated as follows:

\textbf{Client Native (CN)} firstly initiates paging requests and caches the candidate items with corresponding features returned from RS servers. In EdgeRec, the paging size is set to 50, the same as the original RS in Taobao for the stability. While, the number of returned items from RS servers is set to 100 in order to provide more space for reranking on device.\footnote{The configurations are set empirically according to specific RS environment.}
Then, CN collects user behaviors on the exposed items and triggers the model serving module. After receiving the rank of candidate (a.k.a. unexposed) items from model serving module, CN adjusts the UI display for items.


\textbf{Model Serving (MS)} is the core module in EdgeRec system. When triggered by CN, MS begins with the feature engineering for user behaviors and candidate items recived from CN, followed by a neural network based model, with the aims of User Behavior Modeling to capture timely user behaviors and Context-aware Reranking to respond to users in time.\footnote{We utilize MNN (\url{https://github.com/alibaba/MNN}) as our online deep neural network inference engine on device.} Fianlly, MS sends logs to cloud and return the ranking results of candidate items to CN.


\textbf{Recommender System (RS)} on servers can be regarded as a recall module in EdgeRec, which aims to respond to paging requests from CN and provide candidate items with initial ranking. Moreover, it can look up item features and embeddings (e.g., category embedding) which models in MS module need from a key-value storage on cloud for the candidate items before responding to CN.

\textbf{Offline Training (OT)} module first collects logs from MS and constructs samples before model training.
Next, the trained model is split into three parts: (1) sub-model of User Behavior Modeling, (2) sub-model of Context-aware Reranking and (3) embedding matrices (e.g., category and brand). Finally, the first two sub-models are deployed on MS module, while embedding matrices are kept on cloud as the key-value form.



\subsection{System Implementation}
\label{sec:system_implementation}
In this section, we mainly introduce the implementation details of two critical modules in the EdgeRec system:  client native and model serving module. 


\subsubsection{Client Native}
\label{sec:system_implementation_client_native}
One key part of client native is to collect user rich behaviors on client in mobile Taobao RS, e.g. browsing and click records (more detailed behaviors can be seen in Sec. \ref{sec:feature}).
These user behaviors are then stored in a database on device.
As running of EdgeRec models (a.k.a. Model Serving) is triggered by client native,
another critical part is the strategy to trigger it.
Here we set several trigger points according to user's online real-time behaviors, which are summarized as follows: (1) user clicks an item, (2) user deletes an item (i.e. long press) and (3) $k$ items have been exposed without clicked.
We argue that these three types of user behaviors reveal user's current preferences on RS and RS should respond him/her in time (i.e. trigger Model Serving).


\subsubsection{Model Serving}
\label{sec:edgerec_model_serving}
Deep neural network models serving on mobile device face a lot of challenges against traditional cloud serving, e.g., overhead of computing and storage.
There are two critical implementations for EdgeRec model serving
targeting on computing and storage efficiency respectively.
The idea is to \textbf{distribute the model across edge and cloud}, which makes EdgeRec support serving large scale neural network models on mobile device for RS.
\paragraph{\textbf{Computing Efficiency}}
User Behavior Modeling (see Sec. \ref{sec:user_state}) and Context-aware Reranking (see Sec. \ref{sec:rerank}) are trained together but deployed separately and run asynchronously on device.
User Behavior Modeling utilizes recurrent neural network (RNN) \cite{mikolov2010recurrent} based sequence modeling approach which is much inefficient if it always does inference from start (i.e. with $O(n)$ time complexity).
So it is independently inferred along with users' online incoming behaviors in real-time by the recurrent characteristic of RNN (i.e. with $O(1)$ time complexity) and produces behavior encodings which will be stored in database on device.
Context-aware Reranking will first retrieve the behavior encodings from database and then do model inference based on them.
\paragraph{\textbf{Storage Efficiency}}
ID type of features are common and important in RS models, we always utilize techniques of embedding \cite{zhou2018deep,grbovic2018real} to transform them.
However they face challenge of storage efficiency when serving on mobile device.
For example, item brand in our model is an ID feature with about 1.5 million size of dictionary 
and will be transformed into 40 size of hidden state by embedding layer,
in which the embedding matrix will be size of $1500000\times40$ (i.e. about 230MBs).
Models with such large embedding matrices will suffer from overhead of storage when deployed on mobile device.
In our proposed system, we extract embedding matrices from the trained model to deploy in a key-value database on cloud.
And these embedding matrices will be retrieved by corresponding items when RS on servers responds to the paging requests from client native and sent to client as item features.
The rest parts of model without embedding layers serving on device (i.e. the model is about 3MBs) will take the embedding features as inputs and then do model inference.

Furthermore we design a model version strategy to ensure model synchronization when it is updated,
because successfully deploying model on device may be far delayed against deploying model (i.e. the embedding matrices) on cloud,
which depends on the current status of users' mobile device (e.g., connected to wifi or 3G).
In EdgeRec system, we will generate a unique version ID for every trained model. This version ID is kept along with the model deployed on device and embedding matrices stored on cloud. Client native first initiates paging requests with the model version ID on device, then RS on cloud gets the model version ID and retrieves the embedding matrices with corresponding version before responding to the client.
\section{Algorithm}
\label{sec:algorithm}
In this section, we introduce the feature system and methods for user behavior modeling and context-aware reranking. 

\subsection{Problem Definition}
\label{sec:problem}
Our proposed EdgeRec system aims to adapt reranking method on edge targeting waterfall flow recommendation scenario. 
Given the initial ranked item list $S_r$ cached on edge which is generated by the existing RS on cloud, for the reranking request $r\in \mathcal{R}$ in Model Serving module triggered by Client Native module,
our goal is to find a scoring function $\phi (\bm{x}_i, \bm{s}, \bm{C})$ considering (1) features of item $i$ as $\bm{x}_i$, (2) local ranking context from initial model as $\bm{s}$ and 
(3) real-time user behavior context on current recommendation environment as $\bm{C}$.


Reranking models considering local ranking context are well studied in previous works. 
And the local ranking context is represented as the list-wise interaction between initial ranked candidate items which can be modeled by RNN \cite{ai2018learning,zhuang2018globally} or Transformer \cite{pei2019personalized}.
Here we argue that real-time user behavior context is also important for reranking problem especially in waterfall recommendation scenario, while few works consider about it before.
Sec. \ref{sec:user_state} introduces how we model real-time user behavior context with \emph{Heterogeneous User Behavior Sequence Modeling}
and Sec. \ref{sec:rerank} talks about how we model interaction between candidate items and real-time user behavior context with \emph{Context-aware Reranking with Behavior Attention Networks}.
By combining edge computing system and context-aware reranking model, we can achieve \textbf{Real-time Perception} and \textbf{Real-time Feedback} in RS to satisfy users' online varied demands much better.

\subsection{Feature System}
\label{sec:feature}
In this section ,we first give a discussion about our feature system then introduce the detailed user action features for item exposure and item page-view as well as corresponding item feature.
\subsubsection{Insight}
\label{sec:feature_thinking}
In the literature of personalized search and recommender systems~\cite{covington2016deep,zhou2018deep,ni2018perceive,li2019multi,zhou2019deep,pi2019practice,zhou2018micro,Feng:DSIN}, users' behaviors are commonly modeled to characterize their personalized preferences. Hence, these models only consider the direct ``positive feedback'' between users and items (e.g., clicking or transaction), and seldom pay attention to the indirect ``negative feedbacks'' (e.g., skipping or deleting). Although ``positive feedbacks'' are relatively more clear and less noisy, real-time ``negative feedbacks'' are also very important, especially in waterfall flow RS.
Take an intuitive example in online Taobao RS,
after real-time multiple exposures of an item category, 
the corresponding CTR of its re-exposure will decrease significantly.


On the other hand, 
previous works only consider the characteristics (e.g., category and brand) of items interacted with users.
However, user's ``actions''  towards items should also be concerned.
For example, after user clicking an item,
the actions (e.g., adding to favorite and adding to cart) in its detail page (called item page-view) reflect the real user preference towards the item.
Moreover, although user doesn't click an item,
the actions on this item exposure (e.g scroll speed and exposure ) can represent the degree of this item regarded as a ``negative feedback''.
Sometime if a user focuses on an item exposure for a long time without clicking it,
it can't absolutely indicate that he/she doesn't like it.
Especially in current waterfall flow RS, the display of item is getting more and more informative,
e.g., with larger picture, various keywords and even automatically played video,
clicking has become a very ``luxury'' positive feedback for some users.

At last, based on our proposed EdgeRec system,  all the user behavior features are collected, extracted and consumed on edge (i.e. user's mobile device), which is potential to break through the limitations of network latency and bandwidth comparing to current cloud-to-edge based RS system. Therefore,  plentiful and detailed behaviors can be incorporated to infer user preference in a more real-time way. In addition, user's raw behaviors are processed and utilized on his/her own mobile device, which can alleviate user data privacy issues to a certain extent.


To summarize, feature system in our work is novel and promoted 
(1) from ``relying on only positive feedback interactions'' to ``simultaneously paying attention to positive and negative feedback interactions'',
(2) from ``concerning on only interacted items'' to ``considering both interacted items and their corresponding actions'',
and (3) from ``in quasi real-time way'' to ``in ultra real-time way''.

\subsubsection{Item Exposure User Action Feature}
\label{sec:ie_action_feature}
Item Exposure (IE) user actions reveal how user behaves on an item exposure in RS's current display page. 
Fig. \ref{fig:user_feature}(a) illustrates an item exposure in waterfall flow RS in mobile Taobao.
And features for corresponding user actions on it can be classified as (see details in Tab. \ref{tab:user_feature})
(1) item exposure statistics ($e_1\sim e_2$),
(2) user scrolling statistics ($e_3\sim e_5$),
(3) user deleting feedbacks ($e_6$)
and (4) time decay ($e_7$).
Here we represent concatenation of $e_1\sim e_7$ for corresponding item $i$ as item exposure action feature vector $\bm{a}_{IE}^i$.

\subsubsection{Item Page-View User Action Feature}
\label{sec:ipv_action_feature}
\begin{figure}[th]
	\centering
	\includegraphics[angle=0, width=1.0\columnwidth]{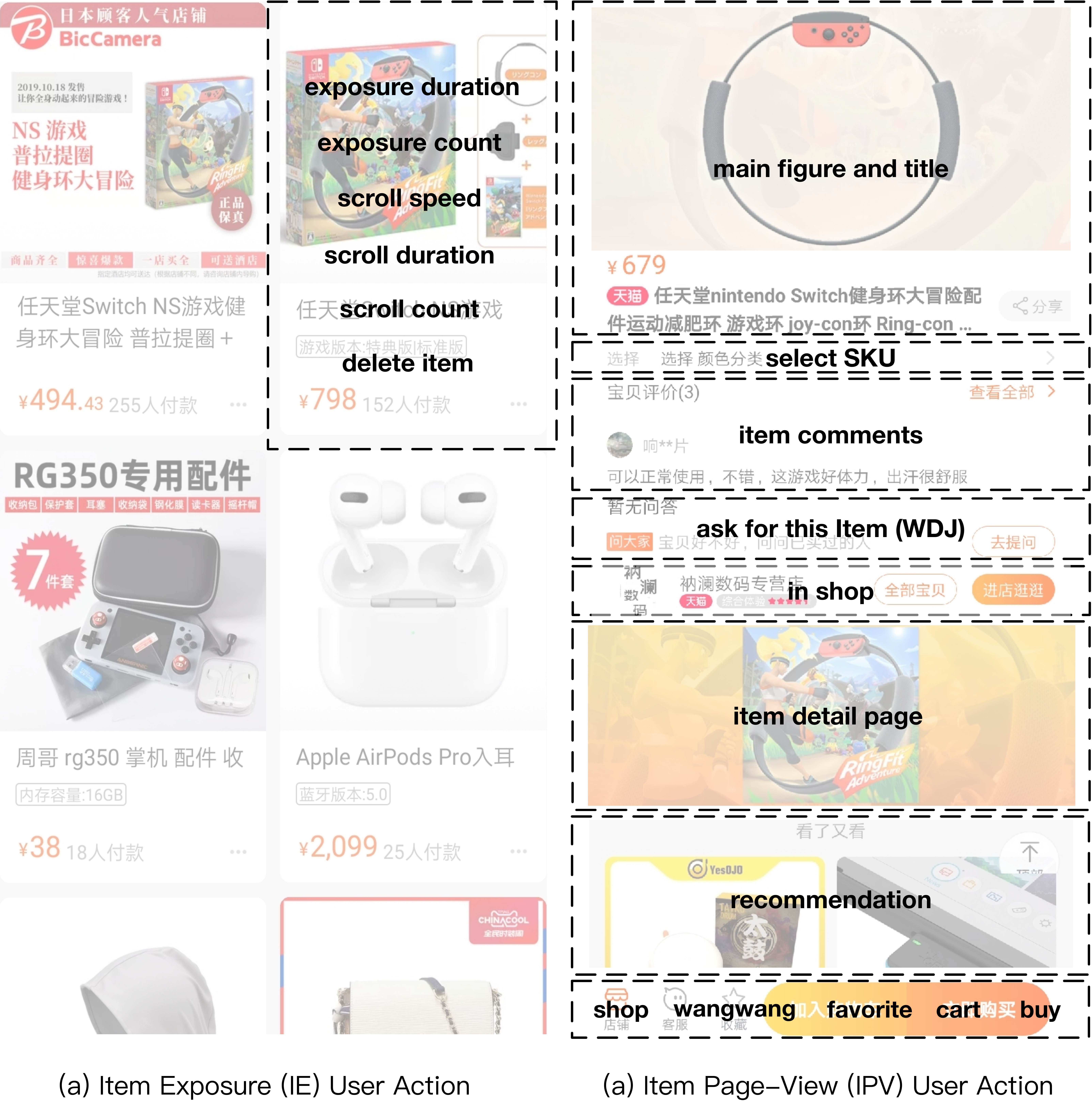}
	\caption{Illustrations for user actions in mobile Taobao.}
	\label{fig:user_feature}
\end{figure}

Item Page-View (IPV) user actions reveal how user behaves in the item detail page
after clicking an item.
Fig. \ref{fig:user_feature}(b) illustrates an item page-view in mobile Taobao.
And features for corresponding user actions in it can be classified as (see details in Tab. \ref{tab:user_feature})
(1) item page-view statistics ($d_1$),
(2) click or not on each block ($d_2\sim d_{11}$)
and (3) time decay ($d_{12}$).
Here we represent concatenation of $d_1\sim d_{12}$ for corresponding item $i$ as item page-view action feature vector $\bm{a}_{IPV}^i$.

\subsubsection{Item Feature}
\label{sec:item_feature}
Apart from the features for user action,
we need features for corresponding item.
And they can be classified as (see details in Tab. \ref{tab:user_feature})
(1) discrete features which are learned by embeddings ($p_1\sim p_6$)
and (2) raw features which are provided from the base ranking models ($p_7$).
Here we represent concatenation of $p_1\sim p_7$ for item $i$ as item feature vector $\bm{p}^i$.
\begin{table}[h]
	\scriptsize
	\caption{Details for the feature system ($e_1\sim e_7$: item exposure user action features; $d_1\sim d_{12}$:  item page-view user action features; $p_1\sim p_7$: item features).}
	\centering
	\begin{tabular}{cccc}
		\toprule
		\bf{Var} & \bf{Attribute} & \bf{Description} & \bf{Type} \\
		\midrule
		$e_1$ & exposure\_duration & total duration of item exposure & bucketize \\
		$e_2$ & exposure\_count & total count of item exposure & bucketize \\
		$e_3$ & scroll\_speed & max scroll speed for item exposure & bucketize \\
		$e_4$ & scroll\_duration & max scroll duration for item exposure & bucketize \\
		$e_5$ & scroll\_count & total scroll count for item exposure & bucketize \\
		$e_6$ & delete\_reason & reason for deleting item (long press) or not & one-hot \\
		$e_7$ & expose\_decay & time decay from item exposure to now & bucketize \\
		\midrule
		$d_1$ & ipv\_duration & total duration in item page-view & bucketize \\
		$d_2$ & cart & add to cart & binarize \\
		$d_3$ & buy & buy immediately & binarize \\
		$d_4$ & favorite & add to favorite & binarize \\
		$d_5$ & comment & go in comment page & binarize \\
		$d_6$ & select\_SKU & select Stock Keeping Unit (SKU) & binarize \\
		$d_7$ & WDJ & ask for this item & binarize \\
		$d_8$ & wangwang & click customer service & binarize \\
		$d_9$ & detail & go in item detail page & binarize \\
		$d_{10}$ & shop & go in shop & binarize \\
		$d_{11}$ & recommendation & go in recommendation page & binarize \\
		$d_{12}$ & ipv\_decay & time decay from item page-view to now & bucketize \\
		\midrule
		$p_1$ & category & product category & embedding \\
		$p_2$ & brand & product brand & embedding \\
		$p_3$ & gender & product suitable gender & embedding \\
		$p_4$ & price\_level & product price level & embedding \\
		$p_5$ & age\_level & product age level & embedding \\
		$p_6$ & bc\_type & product bc type & embedding \\
		$p_7$ & scores & product feature scores (e.g., ctr, cvr, etc.) & raw \\
		\bottomrule
	\end{tabular}
	\label{tab:user_feature}
\end{table}

\begin{figure*}[th]
	\centering
	\includegraphics[angle=0, width=2.0\columnwidth]{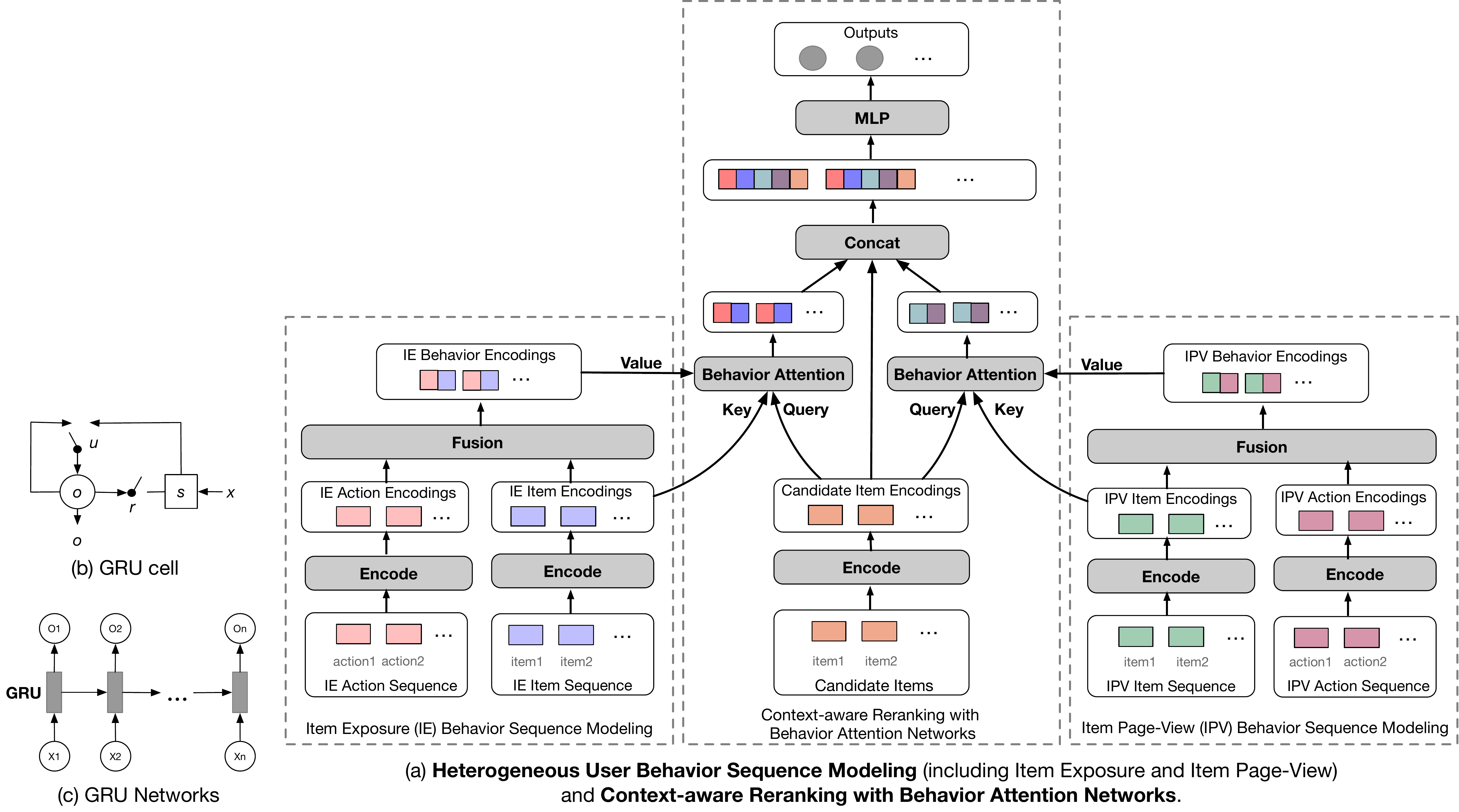}
	\caption{The network architecture of the proposed algorithm in EdgeRec.}
	\label{fig:algo}
\end{figure*}

\subsection{Heterogeneous User Behavior Sequence Modeling}
\label{sec:user_state}
In this section,
we will introduce how to model real-time user behavior context defined as $\bm{C}$.
Following the works \cite{zhou2018deep,ni2018perceive,covington2016deep},
we apply sequence modeling approach as well. 
However previous works only consider user's positive interacted items as we discussed in Sec. \ref{sec:feature_thinking}
so that they cannot handle user behavior sequence modeling well based on our proposed feature system.
Challenge is that there are two aspects of heterogeneity 
in user's behavior data.
In our work, we propose \emph{Heterogeneous User Behavior Sequence Modeling} (HUBSM) particularly targeting on the following two heterogeneities.

The first is heterogeneity of ``item exposure behavior'' and ``item page-view behavior''.
As item clicks are much more sparse compared to item exposures in RS,
if they are encoded together in one sequence, we believe that item page-view behaviors will be dominated.
So we chose to \textbf{model them separately}
(i.e. Item Exposure Behavior Sequence Modeling and Item Page-View Behavior Sequence Modeling).
The second is heterogeneity of ``user behavior actions'' and corresponding ``user interacted items'' which represents two kinds of feature space.
User behavior action features 
reveal the distribution of how user behaves upon an item while item features 
represent the distribution of corresponding item characteristics.
We choose to \textbf{encode them separately first and then do the fusion} for concerning about \emph{Behavior Attention} mechanism in the following context-aware reranking model (see Sec. \ref{sec:rerank}).

Here, the commonly used gate recurrent unit is adopted as our encoder function~\cite{cho2014properties}, controlling the update of network states with an update gate and a reset gate (Fig. \ref{fig:algo}(b)). And we define the sequence encoder function with multi layers GRU networks as follows:

\begin{small}
\begin{eqnarray}
	(\widehat{\bm{X}}, \bm{s}) = \textbf{GRU}(\bm{X}),
\end{eqnarray}
\end{small}
where $\bm{X}=\{\bm{x}^{i}\}_{1\leq i \leq n}$ is input sequence of feature vectors, $\widehat{\bm{X}}=\{\widehat{\bm{x}}^{i}\}_{1\leq i \leq n}$ is output sequence of encodings and $\bm{s}$ is final state of RNN.
The fusion function here is a simple concatenation of two input sequence of feature vectors
$\bm{X}=\{\bm{x}^{i}\}_{1\leq i \leq n}$ and $\bm{Y}=\{\bm{y}^{i}\}_{1\leq i \leq n}$, and is defined as follows:
\begin{small}
	\begin{eqnarray}
	\bm{Z} = \textbf{CONCAT}(\bm{X}, \bm{Y}),
	\end{eqnarray}
\end{small}
where $\bm{Z}=\{\bm{z}^{i}\}_{1\leq i \leq n}$ is the output sequence of fused encodings.
Of course, more sophisticated encoding models (e.g., Transformer~\cite{vaswani2017attention})
and fusion functions (e.g., DNN) can be adopted here.
Considering the size of models on device, we use GRU and concatenation in our implementation respectively.

In the following two paragraphs,
we will formally define our two specific \emph{Item Exposure Behavior Sequence Modeling}
and \emph{Item Page-View Behavior Sequence Modeling} (see Fig. \ref{fig:algo}(a)),
in which user behavior context $\bm{C}$ is represented by two corresponding tuples as $(\widehat{\bm{P}}_{IE},\widehat{\bm{B}}_{IE})$ and $(\widehat{\bm{P}}_{IPV},\widehat{\bm{B}}_{IPV})$.
We deploy HUBSM on device in EdgeRec. 
Based on the recurrent computation characteristics of RNN, we model online incoming user behaviors synchronously and real-time as discussed in Sec. \ref{sec:edgerec_model_serving}.

\paragraph{\textbf{Item Exposure Behavior Sequence Modeling}}
We define IE input sequence of action feature vectors as $\bm{A}_{IE} = \{\bm{a}_{IE}^{i}\}_{1\leq i \leq m}$
and corresponding item feature vectors as $\bm{P}_{IE} = \{\bm{p}_{IE}^i\}_{1\leq i \leq m}$.
Here $m$ is the predefined maximum length of IE behavior sequence and we apply zero padding for shorter sequences.
We get IE output sequence of action encodings $\widehat{\bm{A}}_{IE}$, item encodings $\widehat{\bm{P}}_{IE}$ and fused behavior encodings $\widehat{\bm{B}}_{IE}$ respectively as following equations.
\begin{small}
\begin{eqnarray}
	(\widehat{\bm{A}}_{IE} = \{\widehat{\bm{a}}_{IE}^i\}_{1\leq i \leq m},\ \_) = & \textbf{GRU}(\bm{A}_{IE}), \\
	(\widehat{\bm{P}}_{IE} = \{\widehat{\bm{p}}_{IE}^i\}_{1\leq i \leq m},\ \_) = & \textbf{GRU}(\bm{P}_{IE}), \\
	\widehat{\bm{B}}_{IE} = \{\widehat{\bm{b}}_{IE}^i\}_{1\leq i \leq m} = & \textbf{CONCAT}(\widehat{\bm{A}}_{IE}, \widehat{\bm{P}}_{IE}).
\end{eqnarray}
\end{small}
\vspace{-10pt}

\paragraph{\textbf{Item Page-View Behavior Sequence Modeling}}
We define IPV input sequence of action feature vectors as $\bm{A}_{IPV} = \{\bm{a}_{IPV}^{i}\}_{1\leq i \leq n}$
and corresponding item feature vectors as $\bm{P}_{IPV} = \{\bm{p}_{IPV}^{i}\}_{1\leq i \leq n}$.
Here $n$ is the predefined maximum length of IPV behavior sequence and we apply zero padding for shorter sequences.
We get IPV output sequence of action encodings $\widehat{\bm{A}}_{IPV}$, item encodings $\widehat{\bm{P}}_{IPV}$ and fused behavior encodings $\widehat{\bm{B}}_{IPV}$ respectively as following equations.
\begin{small}
\begin{eqnarray}
	(\widehat{\bm{A}}_{IPV} = \{\widehat{\bm{a}}_{IPV}^i\}_{1\leq i \leq n},\ \_) = & \textbf{GRU}(\bm{A}_{IPV}), \\
	(\widehat{\bm{P}}_{IPV} = \{\widehat{\bm{p}}_{IPV}^i\}_{1\leq i \leq n},\ \_) = & \textbf{GRU}(\bm{P}_{IPV}), \\
	\widehat{\bm{B}}_{IPV} = \{\widehat{\bm{b}}_{IPV}^i\}_{1\leq i \leq n} = & \textbf{CONCAT}(\widehat{\bm{A}}_{IPV}, \widehat{\bm{P}}_{IPV}).
\end{eqnarray}
\end{small}
\vspace{-10pt}

\subsection{Context-aware Reranking with Behavior Attention Networks}
\label{sec:rerank}

In this section, we zoom into details for our reranking approach known as \emph{Context-aware Reranking with Behavior Attention Networks} (see Fig. \ref{fig:algo}(a)) to jointly capture local ranking context and interactions between candidate items and real-time user behaviors context. Following~\cite{ai2018learning}, we encode the sequence of candidate items ranked by initial ranking models with GRU networks and apply the final state as local ranking context $\bm{s}$. With the help of attention technique, our re-ranking model can automatically (soft-)search for parts of user behavior context that are relevant to ranking a target item. Previous CTR prediction models (e.g., DIN~\cite{zhou2018deep} and DUPN~\cite{ni2018perceive}) only learn to attend user historically interacted items with respect to the target item, and thus they fail to model user behavior actions based on above attention mechanism.
As a comparison, our approach  first attend relevant interacted items (a.k.a. find similar interacted items) from user behavior context,
and then attentively combine the corresponding user behavior actions which indicate user's underlying intention to those items,
together represented as context directing the prediction for target item.
We call it \emph{Behavior Attention} here, which specifically employs both item exposure behavior context and item page-view behavior context.

\paragraph{\textbf{Candidate Item Sequence Encoder}}
We define candidate item sequence as $\bm{P}_{CND} = \{\bm{p}_{CND}^{i}\}_{1\leq i \leq k}$,
which is generated and ranked by the prior models in RS servers.
Here $k$ is the predefined maximum length of candidate item sequence and we apply zero padding for shorter sequences.
We apply GRU networks to encode it and represent the final state of RNN as local ranking context in the following equation.
\begin{small}
\begin{eqnarray}
(\widehat{\bm{P}}_{CND} = \{\widehat{\bm{p}}_{CND}^{i}\}_{1\leq i \leq k},\ \bm{s}_{CND}) = & \textbf{GRU}(\bm{P}_{CND}),
\end{eqnarray}
\end{small}
where $\widehat{\bm{P}}_{CND}$ is the output sequence of candidate item encodings
and $\bm{s}_{CND}$ represents the local ranking context.

\paragraph{\textbf{Behavior Attention}}
\label{sec:reranking_target_attention}
Specific for the target candidate item $t$ encoded as $\widehat{\bm{p}}_{CND}^{t}$,
we first attend to the encodings of user behavior item sequence $\widehat{\bm{P}}_{IE}$ and $\widehat{\bm{P}}_{IPV}$ for item exposure and item page-view behaviors respectively.
Then we indicate the attention distributions as $\{att_{IE}^{tj}\}_{1\leq j \leq m}$
and $\{att_{IPV}^{tj}\}_{1\leq j \leq n}$ following Bahdanau attention mechanism~\cite{bahdanau2014neural}.
Finally we produce user behavior contexts known as $\bm{c}_{IE}^t$ and $\bm{c}_{IPV}^t$ 
by combining the attention distributions and fused behavior encodings of user behavior sequence $\widehat{\bm{B}}_{IE}$ and $\widehat{\bm{B}}_{IPV}$.

Specifically following the notations of triplet (Query, Key, Value) in Transformer \cite{vaswani2017attention},
we define $\widehat{\bm{p}}_{CND}^{t}$ as \textbf{Query}, $\widehat{\bm{P}}_{IE}$/$\widehat{\bm{P}}_{IPV}$ as \textbf{Key}, and $\widehat{\bm{B}}_{IE}$/$\widehat{\bm{B}}_{IPV}$ as \textbf{Value} in our model.
We consider that attention calculation here is to (soft-)find similar or relevant items,
so that representations of the compared two feature spaces should be homogeneous.
That's why we choose to encode ``user behavior actions'' and corresponding ``user interacted items'' separately in Sec. \ref{sec:user_state},
and employ user behavior item sequence as \textbf{Key} respect to the \textbf{Query} of target item.
You can see the following equations for details.
\begin{small}
\begin{eqnarray}
\label{eq:attention1}
	& att_{IE}^{tj} = \text{softmax}(v_1^T\text{tanh}(W_1\widehat{\bm{p}}_{CND}^{t} + W_2\widehat{\bm{p}}_{IE}^j)), 1\leq j \leq m, \nonumber\\
	& \bm{c}_{IE}^t = \sum_{j=1}^{m}att_{IE}^{tj}\widehat{\bm{b}}_{IE}^j, \\
\label{eq:attention2}
& att_{IPV}^{tj} = \text{softmax}(v_2^T\text{tanh}(W_3\widehat{\bm{p}}_{CND}^{t} + W_4\widehat{\bm{p}}_{IPV}^j)), 1\leq j \leq n, \nonumber\\
& \bm{c}_{IPV}^t = \sum_{j=1}^{n}att_{IPV}^{tj}\widehat{\bm{b}}_{IPV}^j,
\end{eqnarray}
\end{small}
where weights $W_1$, $W_2$, $W_3$, $W_4$, $v_1$ and $v_2$ are trained parameters.

\paragraph{\textbf{Model Learning}}
To model $\phi(\cdot)$, 
we first simply concatenate  user behavior context on IPV and IE (i.e. $\bm{C}$), representation of target candidate item (i.e.  $\widehat{\bm{p}}_{CND}^{t}$)
and local ranking context(i.e. $\bm{s}$) and feed them into multi-layer perception (\textbf{MLP}) for non-linear transformation. Subsequently, a cross-entropy loss is adopted for model training.


\section{Experiments}
\label{sec:exp}
In this section, we demonstrate the effectiveness of our model on real-world Taobao RS dataset through offline and online evaluation.

\subsection{Offline Evaluation}
\label{sec:offline_eval}


\subsubsection{Dataset}
We collect online logs and corresponding item features (Tab.~\ref{tab:user_feature}) from EdgeRec system in mobile Taobao. In particular, we random sample from the logs in two different days (2019-11-14 and 2019-11-15) and split them into train set (22,072,671 samples) and test set (200,000 samples). Moreover, the average length of IE behavior sequence and IPV behavior sequence of our collected dataset is 56 and 26, respectively.

\subsubsection{Comparing Methods and Evaluation Protocol}
We compare our model with two representative methods, which are widely applied in the industrial applications, namely \textbf{DNN-rank}~\cite{cheng2016wide} and \textbf{DLCM}~\cite{ai2018learning}. To examine the effectiveness of our proposed \emph{Heterogeneous User Behavior Sequence Modeling} (HUBSM) and \emph{Context-aware Reranking with Behavior Attention Networks} (CRBAN), besides our complete  method \textbf{CRBAN+HUBSM(IE\&IPV)},
we prepare four variants of CRBAN: (1) \textbf{CRBAN+HUBSM(IE)} only considering \emph{Item Exposure Behevior Sequence Modeling} (IE-BSM),
(2) \textbf{CRBAN+HUBSM(IPV)} only considering \emph{Item Page-View Behevior Sequence Modeling} (IPV-BSM)
(3) \textbf{CRBAN+HUISM(IE\&IPV)} modeling user behavior context with DIN~\cite{zhou2018deep} instead of HUBSM though IE-BSM and IPV-BSM are both considered.

We train the models with distributed Tensorflow supported by PAI\footnote{\url{https://data.aliyun.com/product/learn}} with the following training settings: batch size = 512, learning rate = 0.005, \#GRU layers = 3, \#GRU hidden units = 32, \#attention hidden units = 32, MLP hidden size = 32, optimizer = ``Aadm''. Note \textbf{DNN-rank} and \textbf{DLCM} only utilize the features on cloud, since they are incapable of capture features on edge (Tab.~\ref{tab:user_feature}).

GAUC~\cite{zhu2017optimized} is a widely used metric for recommendation by averaging AUC~\cite{fawcett2006introduction} over users.
In our paper, we extend GAUC by averaging AUC over Client Native requests $r\in \mathcal{R}$ in EdgeRec system which can be regarded as a reranking session, which is calculated as follows:
\begin{eqnarray}
	GAUC = \frac{\sum_{r\in \mathcal{R}}{\#impression_r\times AUC_r}}{\sum_{r\in \mathcal{R}}{\#impression_r}},
\end{eqnarray}
where $\#impression_r$ and $AUC_r$ are the
number of item impressions and AUC corresponding to the request $r$.

\subsubsection{Result Analysis}
Tab. \ref{tab:main_result} shows the performances of GAUC with respect to different methods.
We can see that DLCM outperforms DNN-rank (row 2 vs. 1) which verifies the effectiveness of involving local ranking context to reranking model.
Furthermore, we can see that all the CRBAN based methods outperform DLCM significantly.
Especially we achieve significant \textbf{2\%} relative improvement of GAUC (row 6 vs. 2) for our complete method CRBAN+HUBSM(IE\&IPV).
This demonstrates the advantage of considering real-time user behavior context in reranking model.
So that how to model user behavior context is what we will focus on in the following discussions.

To verify our proposed \emph{Heterogeneous User Behavior Sequence Modeling} method, we compare HUBSM(IE\&IPV) with HUBSM(IE) (row 6 vs. 3) and HUBSM(IPV) (row 6 vs. 4).
Results show that user behaviors of ``positive feedback'' (known as IPV) and ``negative feedback'' (known as IE) both contribute to modeling user behavior context.
We also find that HUBSM(IPV) outperforms HUBSM(IE) (row 4 vs. 3) which shows that IPV user behaviors can be more important than IE user behaviors.
Finally, results by comparing HUBSM(IE\&IPV) and HUISM(IE\&IPV) (row 6 vs. 5) indicates the promotion by considering both interacted items and their corresponding actions with \emph{Behavior Attention} mechanism.
\begin{table}[h]
	\scriptsize
	\caption{Overall performances.
		Here $^*$ indicates statistical significance improvement compared to the
		baseline (DLCM) measured by t-test at $p$-value of 0.05.
	}
	\centering
	\begin{tabular}{cc|c}
		\toprule
		& \textbf{Method} & \textbf{GAUC} \\
		\midrule
		1 & DNN-rank & 0.62531 \\
		\midrule
		2 & DLCM & 0.63552  \\
		\midrule
		3 & CRBAN+HUBSM(IE) & 0.63818  \\
		\midrule
		4 & CRBAN+HUBSM(IPV) & 0.64039  \\
		\midrule
		5 & CRBAN+HUISM(IE\&IPV) & 0.64283 \\
		\midrule
		6 & CRBAN+HUBSM(IE\&IPV) & $\bm{0.64825}^*$ \\
		\bottomrule
	\end{tabular}
	\label{tab:main_result}
\end{table}

\subsection{Online Performance}
\subsubsection{Online A/B Testing}
\begin{figure}[th]
	\centering
	\includegraphics[angle=0, width=1.0\columnwidth]{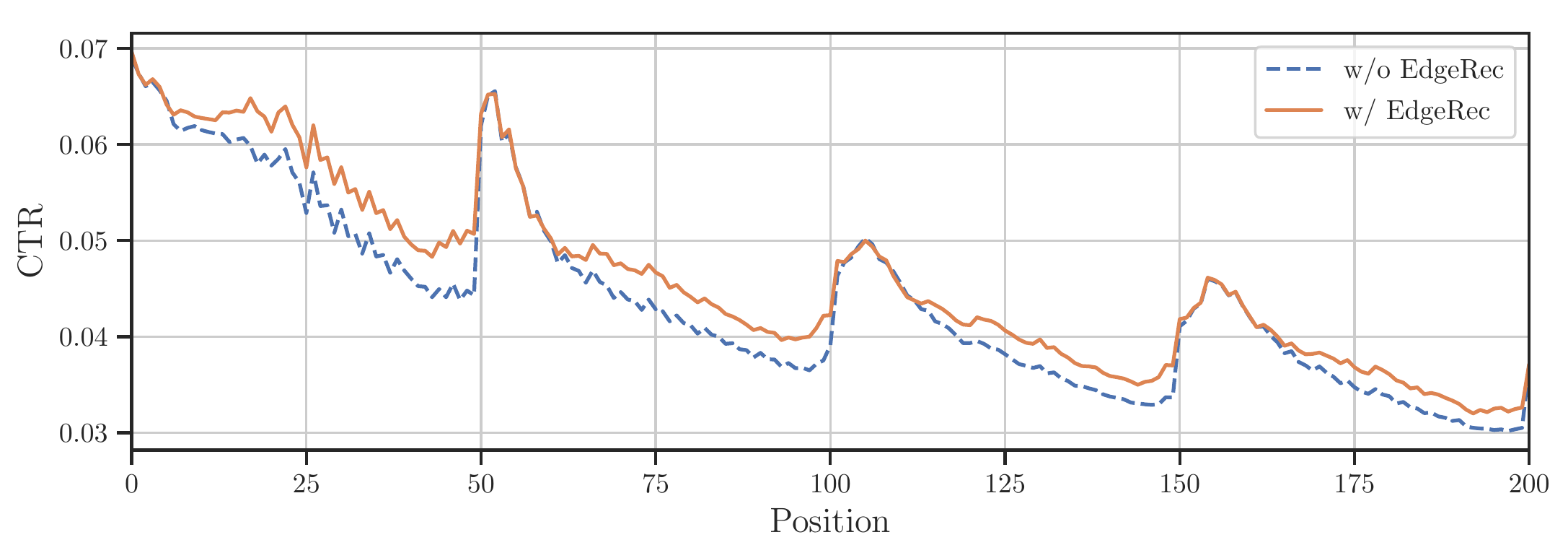}
	\caption{Online averaging item CTR along display position in mobile Taobao RS.}
	\label{fig:pos_ctr_bts}
\end{figure}
We conduct online experiments (a.k.a. A/B testing) on our EdgeRec system deployed in mobile Taobao.\footnote{We utilize a strategy named ``hierarchical bucketing'' in which different algorithms on cloud servers will be transparent for online experiments on edge.}
In Taobao waterfall flow RS, online metrics include PV, CTR, CLICK and GMV,
which evaluate how much willingness for users to view (PV), click (CTR, CLICK) and purchase (GMV) in RS.

EdgeRec has been fully deployed in mobile Taobao application and served billions of users.
The baseline (known as A test) is the conventional Taobao RS without EdgeRec.
Here millions of different random users are involved for online test A and B respectively in the same time.
During almost two weeks' testing from 2019-10-26 to 2019-11-08,
EdgeRec with the complete model CRBAN+HUBSM(IE\&IPV) contributes up to \textbf{1.57\%} PV, \textbf{7.18\%} CTR, \textbf{8.87\%} CLICK and \textbf{10.92\%} GMV promotions on average.
This is a definitely significant improvement and demonstrates the effectiveness of our proposed system.

Furthermore we review the online averaging item CTR along display position in Taobao RS.
Fig. \ref{fig:pos_ctr_bts} shows that CTR at the end of current page will be largely improved after deploying EdgeRec,
which indicates that involving \textbf{Real-time Perception} and \textbf{Real-time Feedback} can greatly increase user click willingness in RS,
as RS is able to satisfy users' online demands in time.


\subsubsection{Online System Performance}
\label{sec:system_performance}
Apart from online A/B testing to show the business performances in production, we also conduct the efficiency of EdgeRec in mobile Taobao
Tab. \ref{tab:system_performance}, revealing the significant improvements of system efficiency from three key aspects after deploying EdgeRec.

\textbf{Delay time for user behaviors} influences the timeliness for system to capture user personalized preference to items which may affect user experience in RS.
Due to the limitations of network bandwidth and latency, RS based on only cloud-to-edge framework may lead to up to 1min delay time for capturing user behaviors.
However after deploying EdgeRec, user behaviors can be collected and consumed on device without any network communication overhead, which can make the delay time within 300ms (e.g., time for reading user behaviors from on-device database).

\textbf{Response time of system} is another factor affecting user experience in RS.
When client native initiates a request to system along user scrolling on RS scenario, the system should respond in time and provide the ranked items to user, otherwise the user will be waiting which may make him/her leave.
Due to the computing overhead of serving such complex RS models for hundreds of millions of users in Taobao,
RS based on only cloud computing may lead to 1s response time including network transmission.
While in EdgeRec, the models are serving on each user's mobile device,
which solves the problem of centralized computing overhead and makes the response time within 100ms without any network communication.

\textbf{Average times of system feedback for users} is an another key factor affecting user experience in RS.
It reflects how often system can adjust the ranking of items which will be  shown to user when he/she browses in RS.
The more frequently the system can adjust results, the more it can satisfy the users' varied demands in RS.
However RS without EdgeRec cannot make the times of system feedback much larger because it will aggravate computing overhead on cloud servers.
So average times of system feedback for users in Taobao RS without EdgeRec is 3 and paging size is 50 in current cloud-to-edge framework (i.e. users make 3 paging requests on average).
In comparison, there is no explicit paging point in EdgeRec but triggered by client native depending on user behaviors (see Sec. \ref{sec:system_implementation_client_native}).
Without increasing additional computing overhead on cloud, the average times of system feedback can be 15 in EdgeRec (i.e. client native triggers 5 reranking requests on average in one page so total is 15).
\begin{table}[h]
	\scriptsize
	\caption{System performances in Taobao RS with (w/) or without (w/o) EdgeRec. They are observed at traffic peak and calculated by averaging over users.}
	\centering
	\begin{tabular}{c|c|c}
		\toprule
		& \textbf{w/ EdgeRec} & \textbf{w/o EdgeRec} \\
		\midrule
		delay time for user behaviors & $\le$ 300ms & $\le$ 1min \\
		\midrule
		response time of system & $\le$ 100ms & $\le$ 1s \\
		\midrule
		times of system feedback & 15 & 3 \\
		\bottomrule
	\end{tabular}
	\label{tab:system_performance}
\end{table}

\subsection{Case Study}
\label{sec:case_study}
\begin{figure}[th]
	\centering
	\includegraphics[width=1.0\linewidth]{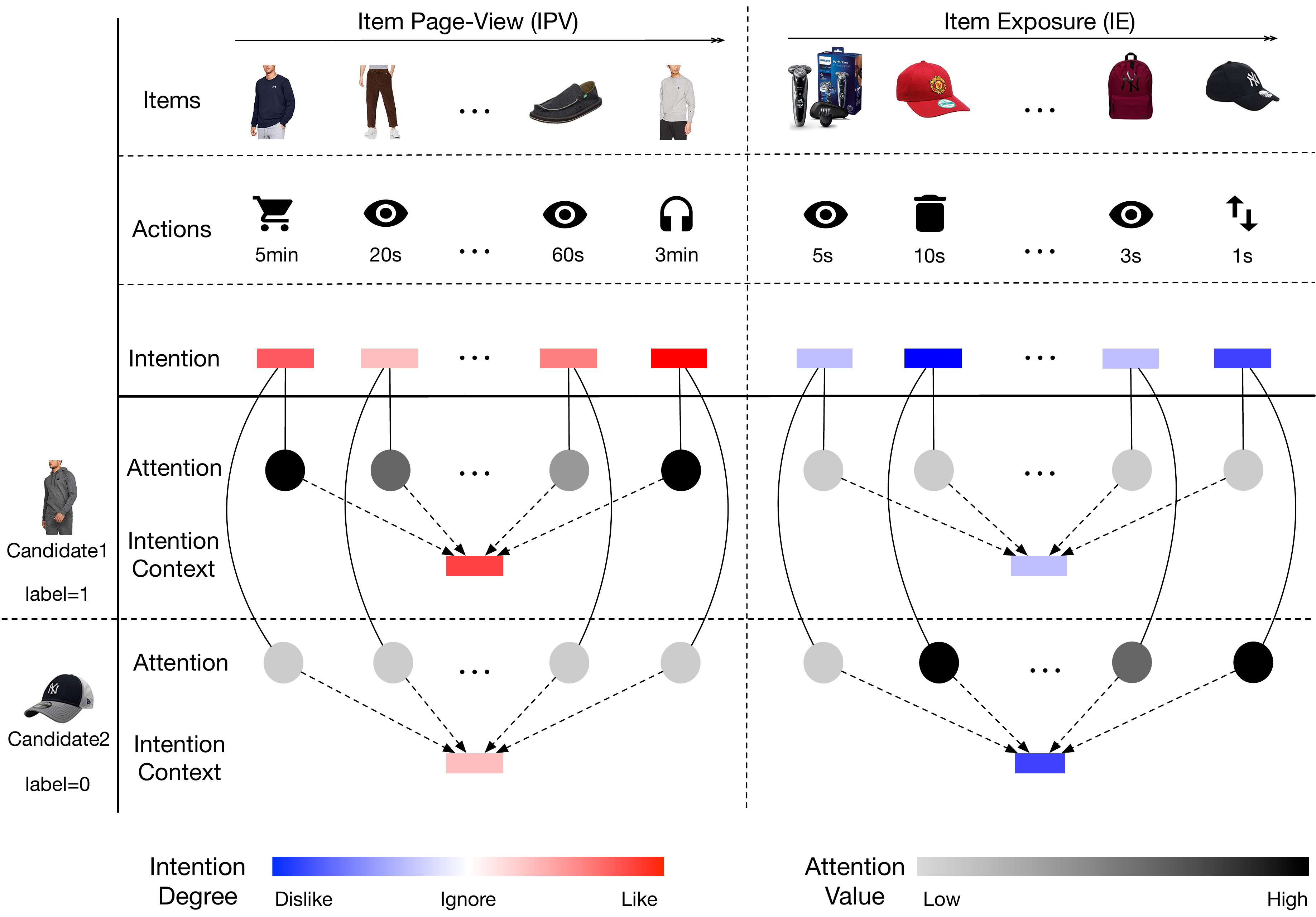}
	\caption{A case in mobile Taobao to illustrate the impact of our proposed HUBSM and CRBAN modules.}
	\label{fig:case_study}
\end{figure}
We conduct a case study on mobile Taobao in Fig. \ref{fig:case_study} to show the effectiveness of HUBSM and CRBAN. In summary, we have the following observations:
(1) user's actions in IPV reveal his/her preference to that item with positive intention degree, e.g., adding to cart or asking for customer service, while user actions in IE usually infer negative intention to that item, e.g., scrolling quickly or deleting. It implies that HUBSM is able to capture user's underlying positive and negative intentions to the historical interacted items.
(2) The candidate shirt is predicted to positive with the help of two similar shirts in IPV, while candidate hat is predicted to negative with two similar interacted items in IE with lower negative intention degrees. It indicats that CRBAN is able to model the interaction between candidate items with user behavior context which directs the prediction for target item better.

\section{Conclusion and Future Work}
\label{sec:conclusion}
We design and implement EdgeRec to tackle the problem of delay for user perception and system feedback in waterfall RS, which is the first attempt to combine RS and edge computing.
Specifically, we propose \emph{Heterogeneous User Behavior Sequence Modeling} and \emph{Context-aware Reranking with Behavior Attention Networks} to model users' plentiful behaviors. Extensive offline and online evaluations verify the effectiveness of EdgeRec in the industrial RS.
We believe that EdgeRec will bring a lot of interesting topics to both industry and research of RS in the future, e.g, thousand people with thousand models (a.k.a. model personalization) based on on-device training, federated learning \cite{hard2018federated,chen2018federated} and interactive recommendation \cite{chen2019large}.



\bibliographystyle{ACM-Reference-Format}
\bibliography{paper}

\end{document}